\documentclass[aip,apl,showpacs,reprint,floatfix,superscriptaddress]{revtex4-1}

\usepackage[utf8]{inputenc}
\usepackage{graphicx}
\usepackage{dcolumn} 
\usepackage{amsmath,amssymb,bm}
\usepackage{esint}
\usepackage[caption=false]{subfig}
\usepackage{color}
\usepackage[]{lineno}

\begin{document}
\title{
Controlling $T_c$ of Iridium films using interfacial proximity effects
}

\author{R. Hennings-Yeomans}
\email{hennings@berkeley.edu}
\affiliation{Department of Physics, University of California, Berkeley, CA 94720 USA}
\affiliation{Nuclear Science Division, Lawrence Berkeley National Laboratory, Berkeley, CA 94720 USA}

\author{C.L. Chang}
\affiliation{High Energy Physics Division, Argonne National
Laboratory, Argonne, IL 60439 USA}
\affiliation{Kavli Institute for Cosmological
Physics, University of Chicago, Chicago, IL 60637 USA}
\affiliation{Department of Astronomy and Astrophysics, University
of Chicago, Chicago, IL 60637 USA}

\author{J. Ding}
\affiliation{Materials Science Division,
Argonne National Laboratory, Argonne, IL 60439 USA}

\author{A. Drobizhev}
\affiliation{Department of Physics, University of California, Berkeley, CA 94720 USA}
\affiliation{Nuclear Science Division, Lawrence Berkeley National
Laboratory, Berkeley, CA 94720 USA}

\author{B.K. Fujikawa}
\affiliation{Nuclear Science Division, Lawrence Berkeley National Laboratory, Berkeley, CA 94720 USA}

\author{S. Han}
\affiliation{Department of Physics, University of California, Berkeley, CA 94720 USA}

\author{G. Karapetrov}
\affiliation{Department of Physics, Drexel University, Philadelphia, PA 19104 USA}

\author{Yu.G. Kolomensky}
\affiliation{Department of Physics, University of California, Berkeley, CA 94720 USA}
\affiliation{Nuclear Science Division, Lawrence Berkeley National Laboratory, Berkeley, CA 94720 USA}

\author{V. Novosad}
\affiliation{Materials Science Division, Argonne National Laboratory, Argonne, IL 60439 USA}
\affiliation{Physics Division, Argonne National Laboratory, Argonne, IL 60439}

\author{T.O'Donnell}
\affiliation{Department of Physics, University of California, Berkeley, CA 94720 USA}
\affiliation{Nuclear Science Division, Lawrence Berkeley National Laboratory, Berkeley, CA 94720 USA}

\author{J.L. Ouellet}
\affiliation{Department of Physics, University of California, Berkeley, CA 94720 USA}
\affiliation{Massachusetts Institute of Technology, Cambridge, MA 02139 USA}

\author{J. Pearson}
\affiliation{Materials Science Division,
Argonne National Laboratory, Argonne, IL 60439 USA}

\author{T. Polakovic}
\affiliation{Department of Physics, Drexel University, Philadelphia, PA 19104 USA}
\affiliation{Physics Division, Argonne National Laboratory, Argonne, IL 60439}

\author{D. Reggio}
\affiliation{Department of Physics, University of California, Berkeley, CA 94720 USA}

\author{B. Schmidt}
\affiliation{Nuclear Science Division, Lawrence Berkeley National Laboratory, Berkeley, CA 94720 USA}

\author{B. Sheff}
\affiliation{Department of Physics, University of California, Berkeley, CA 94720 USA}

\author{R.J. Smith}
\affiliation{Department of Physics, University of California, Berkeley, CA 94720 USA}

\author{G. Wang}
\affiliation{High Energy Physics Division, Argonne National Laboratory, Argonne, IL 60439 USA}

\author{B. Welliver}
\affiliation{Nuclear Science Division, Lawrence Berkeley National
Laboratory, Berkeley, CA 94720 USA}

\author{V.G. Yefremenko}
\affiliation{High Energy Physics Division, Argonne National Laboratory, Argonne, IL 60439 USA}

\date{\today}

\begin{abstract} 
High precision calorimetry using superconducting transition edge sensors requires the 
use of superconducting films with a suitable $T_c$, depending on the application. To advance 
high-precision macrocalorimetry, we require low-$T_c$ films that are easy to fabricate. 
A simple and effective way to suppress $T_c$ of superconducting Iridium through the 
proximity effect is demonstrated by using Ir/Pt bilayers as well as Au/Ir/Au trilayers. 
While Ir/Au films fabricated by applying heat to the substrate during Ir deposition have 
been used in the past for superconducting sensors, we present results of $T_c$ suppression on Iridium 
by deposition at room temperature in Au/Ir/Au trilayers and Ir/Pt bilayers in the range of 
$\sim$20-100~mK.  Measurements of the relative impedance between the Ir/Pt bilayers and 
Au/Ir/Au trilayers fabricated show factor of $\sim$10 higher values 
in the Ir/Pt case.
These new films could play a key role in the development of scalable 
superconducting transition edge sensors that require low-$T_c$ films to minimize heat capacity 
and maximize energy resolution, while keeping high-yield fabrication methods.

\end{abstract}


\maketitle 

\section{Motivation and Background} \label{Sec:Intro}
The first glimpse of physics beyond the Standard Model of particle physics came from the observation of 
neutrino oscillations, which imply that neutrinos have a non-zero, albeit small, mass~\cite{First}. This discovery has 
given emphasis onto the study of the Dirac or Majorana particle nature of the neutrino~\cite{Second}. With two leptons 
in the final state and none in the initial, existence of neutrinoless double-beta decay would 
imply lepton flavor violation by two units and provide a strong clue for theories beyond the Standard Model 
of particle physics~\cite{Third, Fourth, Fifth, Sixth}. An observation of neutrinoless double-beta decay 
would be a direct proof that neutrinos are their own anti-particles and therefore Majorana fermions~\cite{Sixth}. 
Distinction of the energy from the two electrons produced in a neutrinoless double-beta decay, from other 
ordinary processes (backgrounds) requires the state of the art particle detectors that can provide excellent 
energy resolution and timing. 

In the context of improving superconducting Transition Edge Sensors (TES) for 
applications such as next generation searches for neutrinoless double-beta decay~\cite{CUPID}, 
we are investigating how to fabricate superconducting films with a low critical temperature ($T_c$). 
TES technology has been widely used in experiments searching for Dark Matter~\cite{CDMS, CRESST}, 
in measuring the Cosmic Microwave Background~\cite{CMB-1, CMB-2}, and in the detection of x-rays, 
infrared and optical photons~\cite{IrwinHilton}. One of the main features of a TES is its excellent 
energy resolution. When voltage-biased~\cite{KentIrwin}, the energy resolution of the detector is 
proportional to $\sqrt{ T^2 C_{tot}\sqrt{\beta+1}/\alpha}$
where $\alpha \equiv \frac{T}{R}\frac{dR}{dT}$, and $\beta \equiv \frac{I_0}{R_0}\frac{dR}{dI}$.  The total 
heat capacity can be partitioned as $C_{tot} = C_{bolo}(T^3)+C_{TES}(T)+C_{other}$, where 
$C_{bolo}$ is the heat capacity corresponding to the absorber (Debye model), $C_{TES}$ is the term 
associated with the TES material content and $C_{other}$ corresponds to other heat 
capacity contributions, such as those of magnetic impurities in the crystal. There is a 
clear dependence of the energy resolution on the temperature 
hence the need for low-$T_c$ TES in order to maximize the detector sensitivity.

The CUORE Upgrade with Particle IDentification, or CUPID, is a proposed next generation double-beta decay 
experiment that requires a low-threshold optical photon light detector. The use of a secondary bolometer 
as a light detector is being investigated since $\alpha$-particles (the expected main background 
in CUORE~\cite{cuore_background_model}) 
could be tagged by comparing the heat and light signals for each event, given that the two betas from a 
neutrinoless double beta event would produce both heat and light (Cherenkov light in the case of TeO$_2$) 
while $\alpha$ events would produce only heat in the absorber. A baseline resolution of at least 20~eV is 
required~\cite{Casali} in order to achieve a rejection factor of 99.9\% of the $\alpha$ background while 
keeping $>$90$\%$ of the signal events of interest. Furthermore, improvements in timing resolution could 
also be achieved using TES technology.

In order to achieve high production yield of low-$T_c$ TES, we are investigating how to effectively suppress  
$T_c$ using simple fabrication techniques at room temperature. We have studied $T_c$ suppression 
in superconducting Iridium by means of the proximity effect, in which a normal metal is deposited on top (or bottom) 
of an Iridium film. 
Suppression of $T_c$ in the superconducting metal near the superconductor-normal metal interface arises as part of the proximity effect. The Cooper pairs from the superconductor penetrate into the normal metal and induce superconductivity in the metal near the interface (proximity effect) while the normal electrons from the metal diffuse into the superconductor causing suppression of $T_c$ in the superconductor (inverse proximity effect). This phenomenon can be applied to superconducting thin films to tailor their $T_c$ in various applications, including TES first used by 
Nagel et al.~\cite{CRESST-films-1, CRESST-films-2, CRESST-films-2b} and by 
G.~Angloher et al.~\cite{CRESST-films-3} in the context of microcalorimeters for x-ray detection. 
We present two room temperature 
multilayer Iridium depositions, Ir/Pt and Au/Ir/Au. We believe these films could be useful in the development of 
high production yield low-$T_c$ TES in the context of next generation rare-event searches in which large arrays 
of detectors are needed. Furthermore, room temperature deposited films may allow application of TES technology 
not only in semiconductor Si or Ge wafers but also directly on to the bulk of large crystals such as TeO$_2$ or 
Li$_2$MoO$_4$, to name a few examples.

\section{Experimental setup} \label{Sec:FabAndSetup}

\subsection{Film fabrication}
We fabricated superconducting multi-layers by sputtering deposition on high-resistivity ($>$10000 $\Omega$-cm)
silicon wafers, $280\pm10\mu$m thick, at Argonne National Laboratory. The base pressure of the sputtering 
chamber was $10^{-7}$~mbar previous to the 
insertion of the argon gas. The sputtering was performed with Ar gas at 3~mbar, 
allowing for the ignition of the plasma. For each metal target in the chamber, we kept track of the position 
of the target, and all depositions for the same metal were done with the same pressure, conserving the same  
deposition rate. There were two distinct types of sputtering conditions, 
one in which heat was applied during deposition of the iridium layer (followed by deposition of the normal 
metal at room temperature) and the rest in which the iridium base and all the sputter depositions 
were made at room temperature. Deposition rates of about 2.6~\AA/sec for Ir, 2.9~\AA/sec for Au 
and 2.1~\AA/sec for Pt films were used. In the case of Au/Ir/Au trilayers, a 3~nm thick iridium layer was 
deposited prior to the trilayer in order to help the Au/Ir/Au trilayer stick to the silicon wafer. After 
all film layers were deposited, the wafers were diced into squares of 3~mm per side. Subsequently, the 
chips were attached to a copper plate using GE-varnish~\cite{GE-varnish} and wire bonded in 4-wire measurement 
configuration for a resistance measurement. 

\subsection{Resistance measurement and thermometry}

A 4-wire resistance measurement is made using the Lakeshore model 370 AC resistance bridge~\cite{Lakeshore}. 
A 13.7~Hz excitation current is injected on one pair of leads while the voltage is measured on the other 
lead pair. The resistance bridge model 370 applies a phase sensitive detection technique that is 
used in lock-in amplifiers. The uncertainty on the resistance measurement due to the precision of the 
370 AC resistance bridge is between 0.05-0.1$\%$~\cite{Lakeshore}. 
The superconducting multilayers where installed on the mixing chamber plate of an Oxford 
Triton~400 dilution refrigerator unit. 

\begin{figure}
	\includegraphics[width = 3 in]{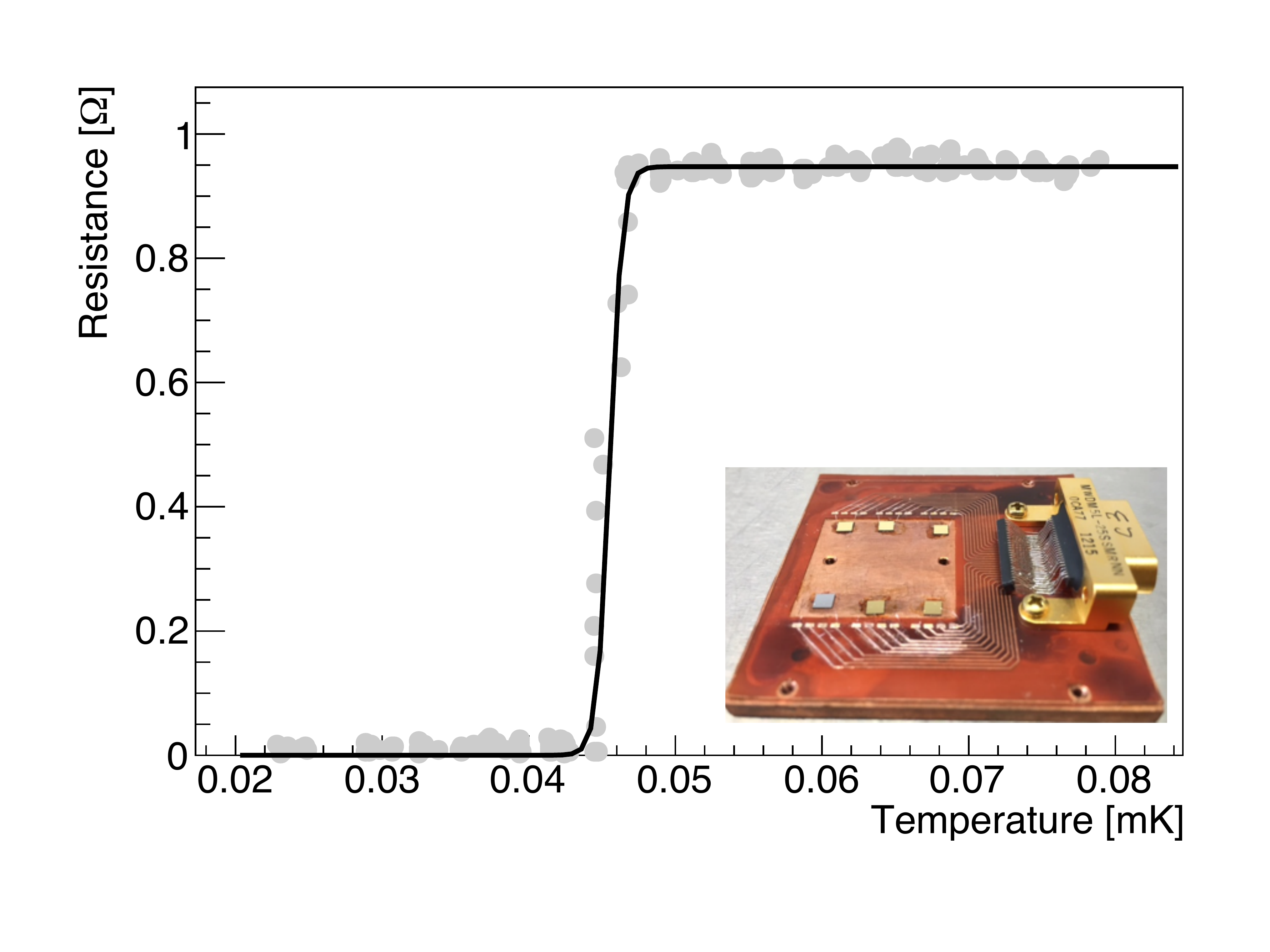}
	\caption{Resistance vs temperature data and fit for an Ir/Pt bilayer where the thickness 
	of Ir=100~nm and Pt=60~nm.  The temperature data is taken using ruthenium oxide thermometer. 
	This bilayer film was sputtered at room temperature and showed a critical temperature 
	T$_c$=46$\pm$1~mK. The insert shows a photograph of the cryogenic mount for six superconducting 
	chips, each bonded for a four-wire measurement of resistance.}
	\label{fig:IrPt_Tc}
\end{figure}

The temperature measurements in the range of 50-200~mK 
were made using a calibrated ruthenium oxide thermometer from Lakeshore cryotronics~\cite{Lakeshore}. 
Between 30-50~mK this thermometer was calibrated against 
a nuclear demagnetization $^{60}$Co thermometer mounted at the center of the mixing chamber plate. 
Between 8-30~mK we utilized either a $^{60}$Co decay anisotropy thermometer or a Johnson 
noise thermometer from Magnicon~\cite{NT-magnicon} that we calibrated against the $^{60}$Co. 
Systematic uncertainty between 30-200~mK is less than 1$\%$ and in the range between 8-30~mK is 
less than 0.5~mK. 

The critical temperature of the superconducting multilayer film samples is determined by a chi-square fit of a function expressed as: 
\begin{eqnarray}
R=D\frac{e^{\left( AT+B\right)}}{1+e^{\left( AT+B\right)}} + C
\label{eq:Tc_fit_fuction}
\end{eqnarray}
where $R$ is the resistance, $T$ is the temperature, $D$ is fixed as the maximum resistance and 
C is a nonzero parasitic resistance. 
The critical temperature is determined by:
\begin{eqnarray}
T_c = -\frac{B}{A}
\label{eq:Tc}
\end{eqnarray}
The dominant systematic error on the $T_c$ measurement is due to the difference in temperature between the mixing 
chamber plate and the superconducting sample during temperature scans across $T_c$. 

Figure~\ref{fig:IrPt_Tc} shows an example of the resistance vs temperature data with a $T_c$ fit for the 
case of an Ir/Pt bilayer in which Ir=100~nm, and Pt=60~nm are the thicknesses of the layers. 

The $T_c$ measurements are made with two excitation currents, 3.16~$\mu$A for films with 
impedances above $\sim$100~m$\Omega$ (Ir/Pt bilayers) and 31.6~$\mu$A for films with 
impedances below it (Au/Ir/Au trilayers) in order to improve signal-to-noise ratio of 
the low impedance films. We performed $T_c$ measurements at different excitation currents 
on one Ir/Pt bilayer (Ir=100~nm, Pt=80~nm) sputtered at room temperature and found a weak 
dependance on the excitation current. Decreasing the bias current from 3.16~$\mu$A to 
0.316~$\mu$A shifted $T_c$ from 20.9$\pm$0.03~mK to 23.0$\pm$0.05~mK. For all $T_c$ 
measurements we report the excitation current used to measure the impedance of the films.

\section{Iridium $T_c$ suppression results} \label{Sec:Results}
\subsection{Ir/Au and Ir/Pt bilayers}

\begin{figure}[]
\begin{center}
\includegraphics[width=3 in]{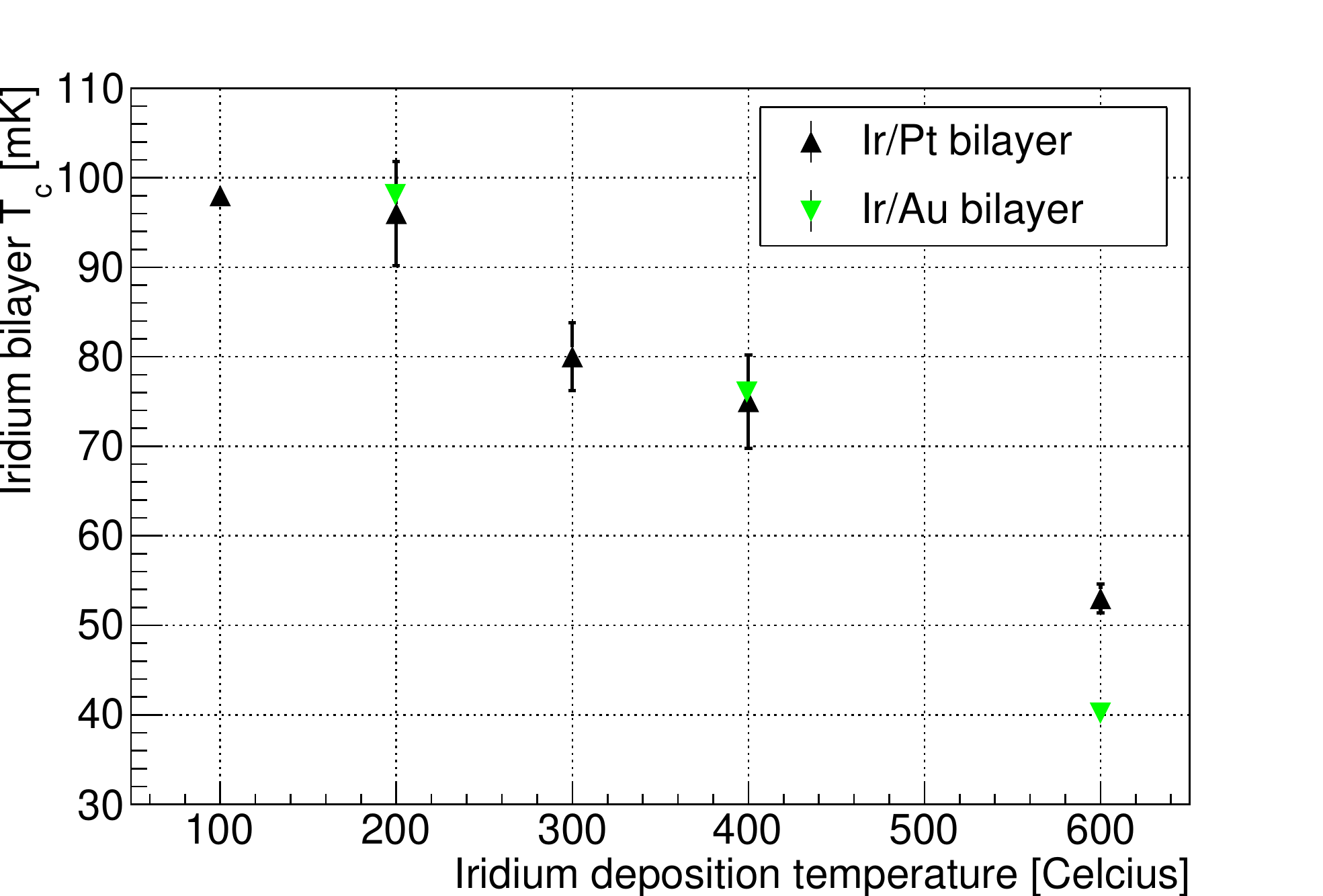}
\caption{
Measurements of $T_c$ of Ir/Pt bilayers (black triangles) and Ir/Au  bilayers (green triangles) in which 80~nm 
iridium base was deposited applying heat to the silicon wafer and then let cool until room temperature before 
deposition of a 20~nm platinum layer and 160~nm gold layer respectively.}
\label{fig:IrPt_IrAu_vs_IrTemp}
\end{center}
\vspace{-20pt}
\end{figure}

We investigated $T_c$ suppression on Ir/Au and Ir/Pt sputtered at room temperature. In principle, 
due to the proximity effect, an increase in normal metal thickness should reduce $T_c$. 
We found this to be true for Ir/Pt all the way to our lowest cryostat temperature ($\sim$8mK). However, 
for the case of Ir/Au bilayer we observed that after a Au thickness of $\sim$200~nm, adding more 
Au would instead increase the $T_c$ of the the Ir/Au bilayer. This effect could be associated 
with the way the films were being fabricated. We moved on to produce Au/Ir/Au trilayers at room 
temperature in which $T_c$ could be suppressed more effectively on both 
sides of the iridium, as shown on Section~\ref{subsec:Au/Ir/Au}.

We also investigated Ir/Au and Ir/Pt films in which heating was applied to the silicon wafer 
during deposition of the iridium base film. Once the iridium deposition is done, we turn off 
the heater and let the sample cool off to room temperature, at which point we proceed to 
sputter the Au or Ir film. We fabricated a set of samples in which only the applied heat during 
iridium deposition was varied and in which the thickness of Ir/Au and Ir/Pt was fixed. Results 
from this set of samples are shown in Figure~\ref{fig:IrPt_IrAu_vs_IrTemp} in which a factor 
of $\sim$2 in $T_c$ suppression is achieved between 200-600$^{\circ}$C for both Ir/Au and Ir/Pt 
bilayer films. 
Suppression of $T_c$ by applying heat during deposition of the iridium 
base layer could be explained if this process increases the quality of the crystaline structure of 
the iridium and also the interface between the superconducting and normal metal.

\begin{figure}[b!]
\begin{center}
\includegraphics[width=3 in]{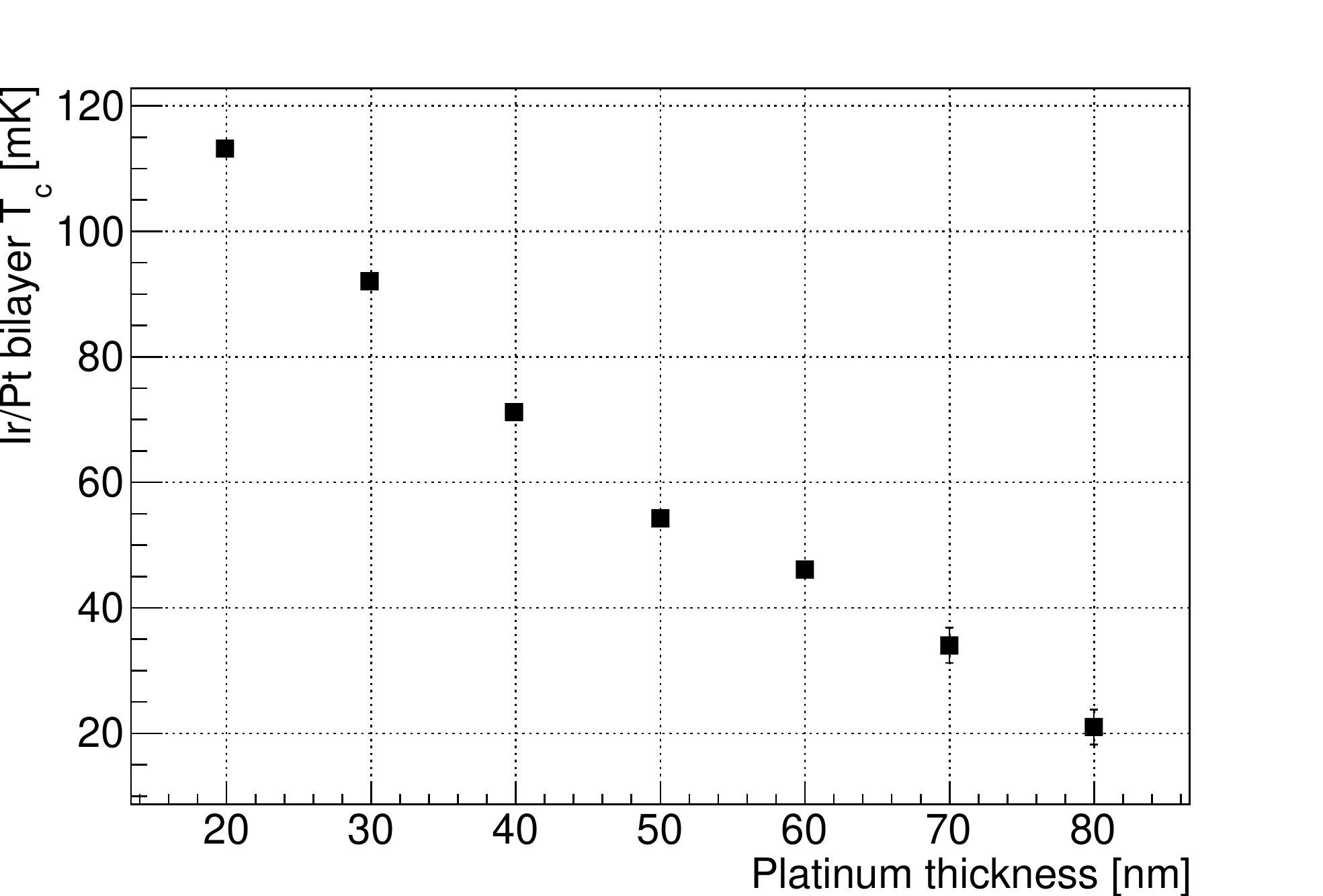}
\includegraphics[width=3 in]{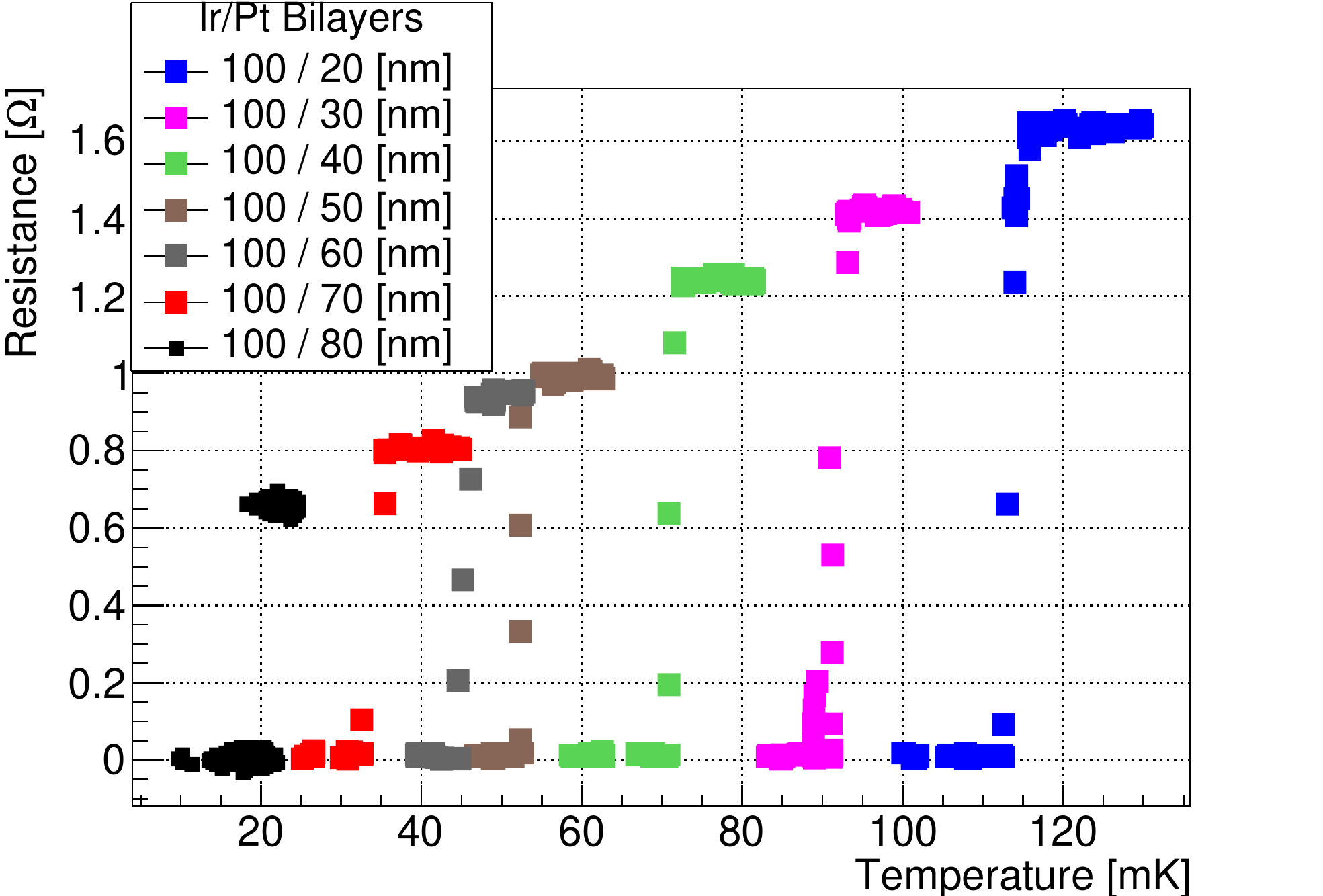}
\caption{
Top) Measurements of the $T_c$ suppression 
in Ir/Pt bilayers in which both the iridium (100~nm) and platinum 
(20-70~nm) layers were deposited at room temperature. A bias current 
of 3.16~$\mu$A was used and the uncertainty in the $T_c$ measurement 
includes the difference between scans increasing or decreasing in 
temperature.  Bottom) Resistance vs temperature data for Ir/Pt bilayers 
from which $T_c$ was obtained. 
}
\label{fig:IrPt_All}
\end{center}
\vspace{-20pt}
\end{figure}

The conditions of the sputtering chamber for Pt depositions were the 
same as for the Ir/Au bilayers, except now a Pt target was used. All 
layers were deposited at room temperature and had a 100~nm iridium 
thickness. The only difference between each film sample was the thickness 
of the Pt layer deposited on top. Results of the $T_c$ measurements are 
shown in Figure~\ref{fig:IrPt_All}. We observe suppression of $T_c$ 
from $\sim$110~mK to $\sim$20~mK for a Pt thickness of 20~nm and 80~nm 
respectively. Figure~\ref{fig:IrPt_All} bottom shows the temperature 
vs resistance curves of the Ir/Pt data shown on top. A bias current 
of 3.16~$\mu$A was used for all the Ir/Pt data. From the above results 
it is clear that Pt has a much stronger pair-breaking effect than Au 
on the Ir superconductor. The suppression of the superconducting temperature 
in S-N bilayers is affected by the density of states near the Fermi energy 
of the two metals and the transparency of the interference barrier~\cite{Martinis}. 
Pt has ten times higher density of electron states at Fermi level than 
gold~\cite{Martin-1, Martin-2} which, assuming the same microstructure of 
Iridium films and barrier transparency, results in approximately eight times 
higher effectiveness of Tc suppression by Pt than by Au as shown in 
Figure~\ref{fig:IrPt_IrAu_vs_IrTemp}.
Therefore, Ir/Pt  is a 
promising room temperature bilayer candidate for low-$T_c$ TES fabrication. 

\subsection{Au/Ir/Au trilayers}
\label{subsec:Au/Ir/Au}
\begin{figure}
\begin{center}
\includegraphics[width=3 in]{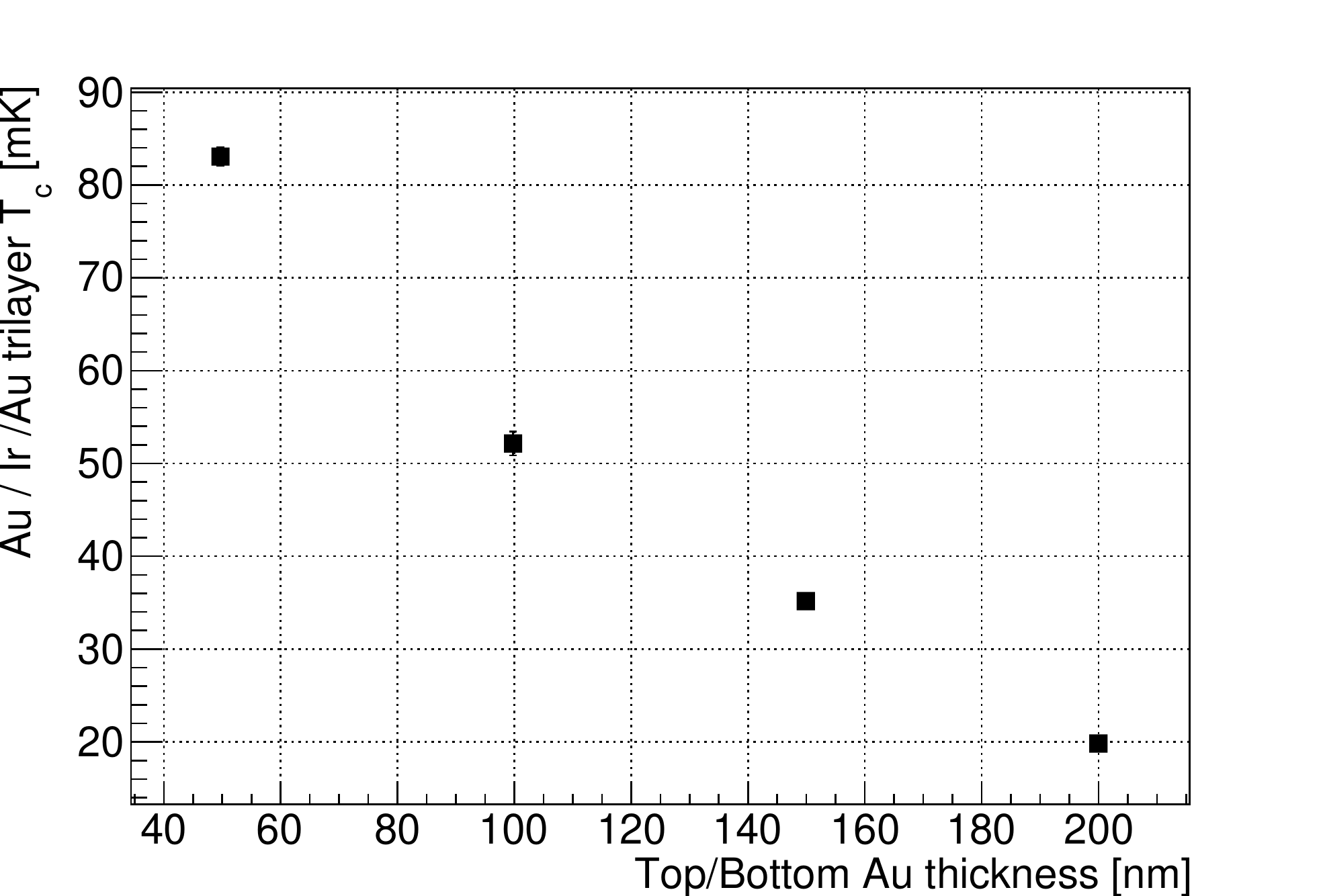}
\includegraphics[width=3 in]{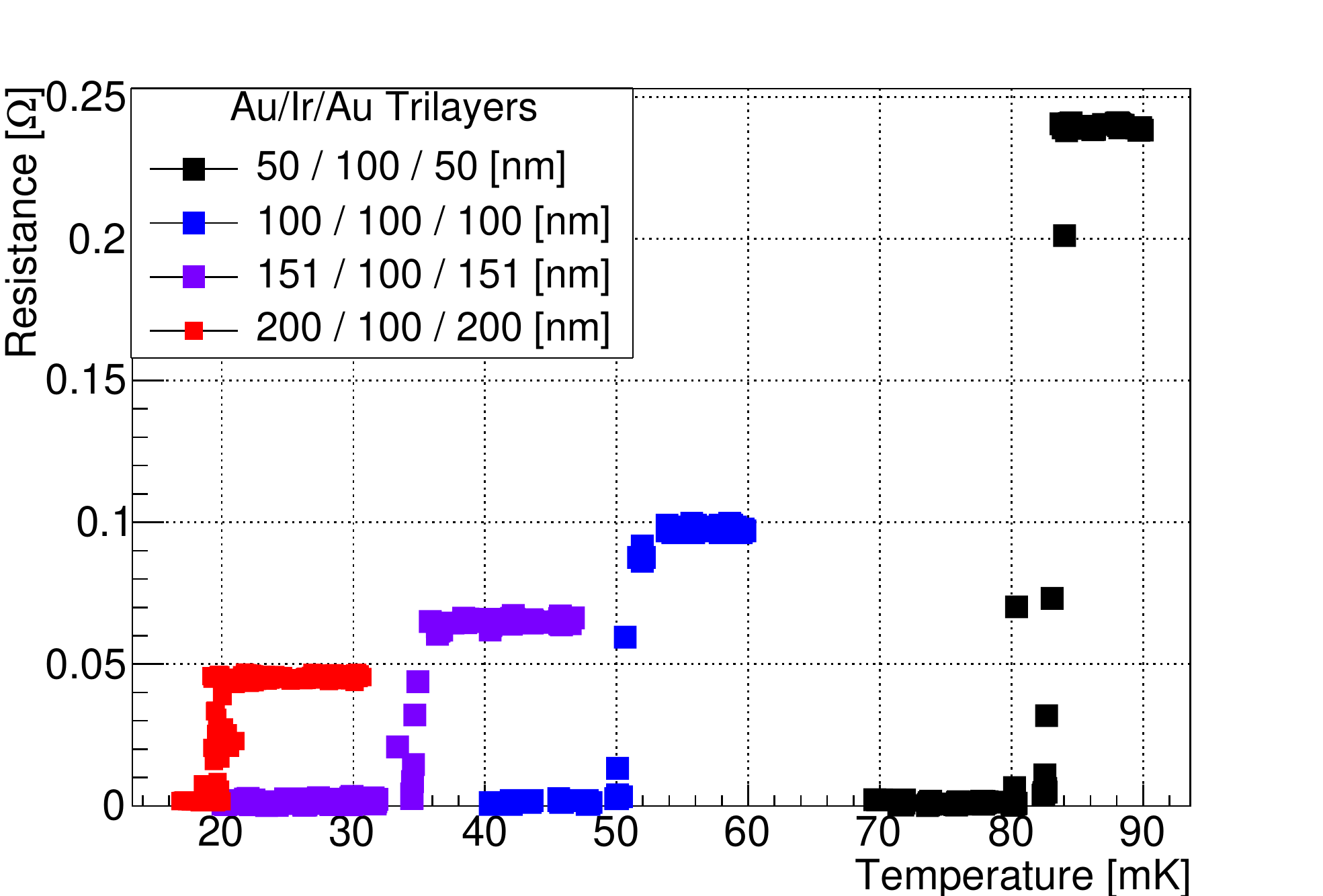}
\caption{
Top) Measured $T_c$ vs gold thickness on top (or bottom) 
for Au/Ir/Au trilayers in which all three layers were deposited at room temperature. 
The iridium thickness is 100~nm for all samples. The same amount of gold was 
deposited on top and bottom of the iridium.  Bottom) Resistance vs temperature data for 
each Au/Ir/Au trilayer. A bias current of 31.6~$\mu$A was used. 
}
\label{fig:Trilayers_all}
\end{center}
\vspace{-20pt}
\end{figure}

Motivated by the goal to have a room temperature fabrication using iridium and gold for low-$T_c$ films, 
we studied $T_c$ suppression using gold above and below a 100~nm iridium film. 
Using the same sputter deposition method, we 
investigated depositing a thinner gold layer on top and bottom of an iridium film. 
We fabricated Au/Ir/Au trilayers at room temperature, all of them with an iridium thickness of 100~nm. The 
gold thickness on top and bottom of the iridium was different for each trilayer.  The results of 
the $T_c$ measurements of these trilayers are shown in Figure~\ref{fig:Trilayers_all}. $T_c$ suppression 
between $\sim$85~mK and $\sim$20~mK is achieved for a gold thickness of 100~nm and 400~nm respectively 
(top+bottom layers) for a bias current of $\sim$31.6~$\mu$A. 

Temperature vs resistance data is shown on Figure~\ref{fig:Trilayers_all} bottom. Ir/Pt bilayers 
show a normal resistance $\sim$10 times higher than the Au/Ir/Au trilayers for a similar $T_c$.  

\section{Conclusion} \label{Sec:Conclusion}

We present $T_c$ suppression between $\sim20-100$~mK of Ir/Pt bilayers and Au/Ir/Au trilayers 
fabricated by sputtering deposition without heating the substrate. The fact that these films 
can be deposited at room temperature allows for the possibility to sputter the TES directly 
on the bulk of the crystals being used as a macrocalorimeter. We also show results of 
$T_c$ suppression in Ir/Au and Ir/Pt bilayers while applying heat to the substrate during deposition 
of the iridium. We believe these measurements to be key in triggering the development 
of superconducting low-$T_c$ TES sensors for large macrocalorimeter arrays such as those utilized 
in experiments searching for Dark Matter or neutrinoless double beta decay.

\begin{acknowledgments}

We would like to thank Paul Barton and Jeff Beeman for help dicing some of the samples and J.G. Wallig for engineering support. This work was supported by the US Department of Energy (DOE) Office of Science under Contract Nos. DE-AC02-05CH11231 and DE-AC02-06CH11357, by the DOE Office of Science, Office of Nuclear Physics under Contract No. DE-FG02-08ER41551, and by the National Science Foundation under grants PHY-0902171 and PHY-1314881. The United States Government retains and the publisher, by accepting the article for publication, acknowledges that the United States Government retains a non-exclusive, paid-up, irrevocable, world-wide license to publish or re- produce the published form of this manuscript, or allow others to do so, for United States Government purposes.

\end{acknowledgments}

\end{document}